\documentstyle[preprint,aps,prb]{revtex}
\gdef\journal#1, #2, #3, 1#4#5#6{
    {\sl #1~}{\bf #2}, #3 (1#4#5#6)}

\def\bR{{\bf R}}

\def\br{{\bf r}}
\def\bh{{\bf h}}
\def\bG{{\bf G}}

\def\eps{\epsilon}

\begin{document}
\draft
\title{\bf Comparison of two methods for describing the strain
profiles in quantum dots}

\author{C. Pryor}
\address{Department of Solid State Physics, Lund University, Lund, Sweden}
\author{J. Kim, L.~W.~ Wang, A.~Williamson  and A.~Zunger}
\address{National Renewable Energy
Laboratory, Golden Co 80401}
\maketitle
\begin{abstract}
The electronic structure of interfaces between lattice-mismatched
semiconductor is sensitive to the strain.  We compare two approaches
for calculating such inhomogeneous strain -- continuum elasticity (CE,
treated as a finite difference problem) and atomistic elasticity (AE).
While for {\it small} strain the two methods must agree, for the large
strains that exist between lattice-mismatched III-V semiconductors
(e.g., 7 \%\ for InAs/GaAs outside the linearity regime of CE) there are
discrepancies.  We compare the strain profile obtained by both
approaches (including the approximation of the correct $C_2$ symmetry
by the $C_4$ symmetry in the CE method), when applied to
$C_2$-symmetric InAs pyramidal dots capped by GaAs.

\end{abstract}
\newpage
\section{Introduction}

One of the leading methods for growing semiconductor
quantum dots is via the controlled coarsening of a film
of a material that is strained with respect to the substrate on which
it is grown.\cite{revSAQD,bimberg95:1}
This (``self-assembled'') coarsening/roughening is
a result of lattice-mismatch-induced strains.
The dots are often capped by the substrate material,
thus extending the strain around the dot to all angular directions.
Not surprisingly, the interpretation of the
electronic structure of such dots
is profoundly affected by their strain profile.
Thus, in order to calculate or interpret
the measured electronic structure,
one has first to calculate or measure the position dependent strain
tensor $\epsilon_{\alpha\beta}$.

The three basic approaches to calculating such strains are:

{\it (i) Harmonic Continuum Elasticity}:
Here, one uses
classical elasticity\cite{landau}
within the harmonic approximation.
For a cubic system, the strain energy per atom, $E_{CE}$, is
\begin{equation}
E_{CE} = {V\over 2}C_{11}
(\epsilon_{xx}^2+\epsilon_{yy}^2+\epsilon_{zz}^2)
+{V\over 2}C_{44}(\epsilon_{yz}^2+\epsilon_{zx}^2+\epsilon_{xy}^2)
+VC_{12}(\epsilon_{yy}\epsilon_{zz}+\epsilon_{zz}\epsilon_{xx}
+\epsilon_{xx}\epsilon_{yy}),
\label {ECE}
\end{equation}
where $V$ is the equilibrium volume, $C_{ij}$ are cubic elastic
constants and $\epsilon_{\alpha\beta}$ is the strain tensor.
We illustrate the predictions of harmonic continuum elasticity
for a 2D film, since this is
going to be used as a test case.
In the absence of
shear strain
($\epsilon_{\alpha\beta} \propto \delta_{\alpha\beta}$),
for a film coherently grown on a substrate with parallel lattice
constant $a_s$, the strain components are
\begin{eqnarray}
\epsilon_{||} = \epsilon_{xx}=\epsilon_{yy} & = &
{a_s-a_{eq}\over a_{eq}}\nonumber\\
\epsilon_{\perp} = \epsilon_{zz} & = & {c-a_{eq}\over a_{eq}},
\label {ST}
\end{eqnarray}
where $a_{eq}$ is the equilibrium lattice constant of the unstrained
material and $c$ is the perpendicular lattice constant of
the strained film.
The equilibrium value of this $c-$axis is determined from
$\partial E_{CE}/\partial \epsilon_{\perp} = 0$,
yielding,
\begin{equation}
{c_{eq}(a_s,{\bf G})\over a_{eq}}-1 = [2-3q({\bf G})]\;\epsilon_{||}(a_s),
\label {EConst}
\end{equation}
where the ``epitaxial strain reduction factor'' for orientation
{\bf G} of the $c-$axis is
\begin{equation}
q({\bf G}) =1-{B\over C_{11}+\gamma({\bf G})\Delta}\;\,
\label {RF}
\end{equation}
and $\Delta = C_{44}-1/2(C_{11}-C_{12})$
is the elastic anisotropy,
$B = 2/3(C_{11}+2C_{12})$ is the bulk modulus
and $\gamma({\bf G})$ is a purely geometric factor given in
Ref.[\cite{alex:elasticity}].
For principal directions,
$\gamma(001) = 0, \gamma(011) = 1$ and $\gamma(111) = 4/3$.
Equations (\ref{ST})--(\ref{RF}) are used routinely to
predict tetragonal distortions of strained films~\cite{alex:elasticity}.
The
 harmonic continuum elasticity method has been
recently applied to pyramidal quantum dots by
Grundmann {\it et al.}\cite{bimberg95:1}
and by Pryor {\it et al.}\cite{pryor}

{\it (ii) Atomistic elasticity}: Here, one avoids a continuum
description and describes the strain {\it energy} in terms of few-body
potentials between actual atoms
\begin{equation}
E_{AE} = \sum_{ij}V_2({\bf R}_i-{\bf R}_{j})
+\sum_{ijk}V_3(\hat\Theta_{ijk})+\cdots,
\label {EAE}
\end{equation}
where $V_2$ is a two-body term, $V_3$ is a three-body function of the
bond angle, $\hat\Theta_{ijk}$.  The functional form of these terms is
taken to be strain-independent.  The strain is determined by
minimizing $E_{AE}$ with respect to atomic positions $\{{\bf R}\}$.
Like the continuum elasticity approach, only the cubic elastic
constants are used as input\cite{vff0,vff1,martins}.  However, unlike
the CE approach, here (a) optical phonon modes can be
described,\cite{vff0,vff1} (b) harmonicity is not assumed, and (c) the
atomic level symmetry is retained.  The last point is illustrated in
Fig.~1 that shows a regular pyramid representing the quantum dots from
experiments of Grundmann {\it et al.}~\cite{bimberg95:1} In a
continuum representation, the strain is equal on the $\{110\}$ and
$\{\bar1 10\}$ facets, while in an atomistic description these two
facets can have different strain if the pyramid is made of a zincblend
material.  Atomistic elasticity has been widely used to determine
strain in alloys,\cite{martins,bellaiche:1}
superlattices\cite{bernard} and dots,\cite{jaros96,jk:1} where $V_2$
and $V_3$ of the Eq.~(5) are taken from Keating's valence force
field~\cite{vff0,vff1,martins} (VFF) model.

{\it (iii) Atomistic quantum-mechanical approach}: Here one does not
have to assume any model for inter-atomic interactions as in the atomic
elasticity.  Instead, one explicitly computes the total electron and
nuclear energy $E_{tot}[\{ {\bf R}_i\}]$ for each atomic configuration
$\{ {\bf R}_i\}$ directly from a quantum-mechanical Schr\"odinger
equation.  Atomic symmetry is retained and harmonicity is not assumed.
This approach has been used for {\it small} ($\leq 100$ atom) wires
and clusters~\cite{buda1,buda2} but it impractical for $\sim 100$ \AA\
dots ($\sim 10^5$ atoms).

The three approaches to the calculation of strain -- harmonic
continuum elasticity, (anharmonic) atomistic elasticity, and the
atomistic quantum mechanical approached -- have been recently compared
for InAs/GaAs strained superlattices.\cite{bernard} However, no
comparison exists for 0-dimensional quantum dots.  Here we perform
parallel calculations for the strain $\widetilde \epsilon ({\bf R})$
of a pyramidal InAs dots (Fig.~1) surrounded by GaAs using the two
approaches that are practical for large dots: continuum elasticity and
atomistic elasticity.  We find that: (i) the strain profiles obtained
via continuum elasticity are in qualitative agreements with those
found by atomistic elasticity; (ii) the atomistic elasticity produces
different strains on the two facets ($\{110\}$ or $\{\bar110\}$) of
the Zincblend pyramidal dots (see Fig.~1), corresponding to the
physical c$_2$ symmetry, while the continuum elasticity approximates
this as c$_4$ symmetric strain.  (iii) the quantitative discrepancy
resides mostly inside the dots, while the difference in the barrier
region is smaller; These differences are traced back to the fact that
the strain lies outside the domain of validity for the linear
elasticity.  We illustrate this point by contrasting the predicted
$c_{eq}(a_s,{\bf G})/a_{eq}$ ratio of coherent 2D films, as obtained
by harmonic continuum elasticity [Eq.~(3)] and atomistic elasticity.
Differences are noticeable already for 1 \% biaxial strain, whereas
the controlled-coarsening (``self-assembled'') growth method for
quantum dots needs to deal with larger mismatches (7 \% for InAs/GaAs
and InP/GaP).  Finally, the consequences on the electronic structure
of the different strain profiles obtained for dots using CE and AE are
illustrated.

\section{Methods of calculations}
\subsection{Continuum elasticity for dots}

In the CE approximation the strain is determined by minimizing the
 elastic energy given in Eq. 1.  To account for the lattice mismatch
 we assume the coordinates are fixed to the barrier material, and
 treat the island as expanded barrier material (with different elastic
 constants). This is accomplished by the modification
\begin{eqnarray}
E_{CE} &\rightarrow& E_{CE}-\alpha({\bf r}) ( \epsilon_{xx} + \epsilon_{yy}
+ \epsilon_{zz})\\
\alpha( {\bf r} ) &=&\cases{
0 & {\rm barrier}\cr
 (C_{xxxx}+2C_{xxyy})(a_I-a_B)/a_B& {\rm island}
}
\end{eqnarray}
where $a_B$ and $a_I$ are the  unstrained lattice constants for the barrier  
and island
material respectively, and $C_{xxxx} $  and $C_{xxyy}$ are the elastic
constants for the island material.
A piece of island material with no external forces acting on it will have its
energy minimum shifted to
\begin{eqnarray}
\epsilon_{ij}= \delta_{ij}(a_I-a_B)/a_B.
\end{eqnarray}
This fictitious strain corresponds to unstrained island material and must
be subtracted. The corrected strain is still computed with derivatives
in the barrier's coordinates and must be converted to the the island
coordinates through multiplication by $dx_b \over dx_I$. Thus, the
physical strain is given by
\begin{eqnarray}
\epsilon_{ij}^{phys}={a_B\over a_I}(\epsilon_{ij} - \delta_{ij}(a_I-a_b)/a_B) 
\label{correction}
\end{eqnarray}
where $\epsilon_{ij}=({du_i\over dx_j}+{du_j\over dx_i})/2$ is the
strain computed directly from the displacement $u_i$ which minimizes
$E_{CE}$.

A numerical solution requires some kind of discretization.  We define
the displacements $u_i$ on a cubic grid, thereby maintaining the cubic
symmetry of the crystal.  The strain is expressed in terms of forwards
or backwards differences by
\begin{eqnarray}
\epsilon^\pm_{ij}= (\Delta^{\pm}_i  u_j  + \Delta^{\pm}_j  u_i  )/2  \\
\Delta_i^+ f({\bf r}) = { f({\bf r+\hat n}_i)-f({\bf r})\over |  {\bf \hat  
n}_i   |}\\
\Delta_i^- f({\bf r}) = { f({\bf r})-f({\bf r-\hat n}_i)\over | {\bf \hat n}_i |}
\end{eqnarray}
where $  {\bf \hat n}_i $ is the lattice vector in the $i$ direction.
Symmetric differences
\begin{eqnarray}
\Delta^s_i f({\bf r}) = { f({\bf r+\hat n}_i)-   f({\bf r-\hat n}_i)\over  
2|  {\bf \hat n}_i   |}\\
\end{eqnarray}
 are undesirable since they give unphysical low energy configurations
which oscillate with period $2 | {\bf \hat n}_i |$.  For example, a
displacement $u_x{(\bf r})=\sin(\pi x/| {\bf \hat n}_x|)$ has
$\epsilon_{xx}=0$ when constructed using symmetric differences.  The
oscillatory solutions cannot be simply discarded since they mix with
the physical ones.  Non-symmetric derivatives are also problematic
since a particular choice will single out a direction in space. The
solution is to average $E_{CE}$ over all permutation of $\pm$ on each
of the three difference operators.  That is, we take
$(E^{+++}+E^{-++}+E^{+-+}+...)/8$.  Physically this corresponds to
taking the energy density at each site to be the average of the energy
densities from each adjoining octant.

The elastic energy is a quadratic function of the displacements, which
is easily minimized using the conjugate gradient algorithm. For the
barrier material the strains are computed directly using differences
(now there is no impediment to using symmetric differences).  In the
island material we then apply the correction in Eq.~(\ref{correction}).

\subsection{Atomistic valence force field for dots}
In the VFF model,
the strain energy
is expressed as a functional of atomic positions, $\{\bR_i\}$, as
\begin{eqnarray}
E_{AE}& = & \sum_{ij}V_2(\bR_i-\bR_j)
+ \sum_{ijk}V_3(\hat\theta_{ijk})\nonumber\\
&=&{1\over 2} \sum_i\sum_j^{nn}
{3\alpha_{ij}\over 8 (d_{ij}^0)^2}
[(\bR_i-\bR_j)^2-(d_{ij}^0)^2]^2\nonumber\\
&&+{1\over 2}\sum_i \sum_{j,k>j}^{nn}
{3\beta_{i,jk}\over 8d_{ij}^0d_{ij}^0}
[(\bR_j-\bR_i)\cdot(\bR_k-\bR_i)-\cos{\theta_0}d_{ij}^0d_{ij}^0]^2.
\label {vffE}
\end{eqnarray}
Here, $d_{ij}^0$ denotes the ideal bond length between atoms $i$ and
$j$, and $\theta_0$ is the ideal bond angle.  For the Zincblend
structure, $\cos{\theta_0} = -1/3$.  The local-environment-dependent
coefficients, $\alpha_{ij}$ and $\beta_{i,jk}$, are fitted to the
elastic constants of bulk materials\cite{vff1}.  The long-range
Coulomb interactions of Ref.~[\cite{vff1}] are neglected which causes
a slight deviation from the measured bulk properties \cite{martins}.
In this case, the elastic constants of a pure bulk Zincblend material
are given as
\begin{eqnarray}
C_{11}+2C_{12} &=& {\sqrt{3}\over 4r}(3\alpha+\beta)\nonumber\\
C_{11}-C_{12} &=& {\sqrt{3}\over r}\beta\nonumber\\
C_{44} &=& {\sqrt{3}\over 4r} {4\alpha\beta\over \alpha+\beta}
\;\;\; ,
\label{econst}
\end{eqnarray}
where $r$ is interatomic bond length.  Because Eq.~(\ref{econst})
contains only two free parameters, it is impossible to fit three
arbitrary elastic constants. Nonetheless, for zincblend materials
$\alpha$ and $\beta$ may be chosen so the $C$'s fit within a few
percent of the measured values.  Table I gives the elastic constants
of bulk GaAs and InAs calculated from Eq.~(\ref{econst}) using
$\alpha$'s and $\beta$'s of Ref.~[\cite{vff1}].  The elastic constants
obtained differ a bit from the experimental values (since Coulomb
corrections to Eq.~(\ref{econst}) are neglected), but we will use them
consistently in both our continuum elasticity and atomistic elasticity
studies. For purposes of comparison, Table I also contains the
experimental elastic parameters.  Note that the VFF method with the
standard parameterization of Eq.(16) reproduces well the values and
trends in the formation enthalpies of strained GaP/InP structures, as
obtained from first-principles\cite{silverman}.

The relaxed atomic configuration is obtained by conjugate gradient
minimization~\cite{cgref} of $E_{AE}$ with respect to the atomic
positions.  At each minimization step, the atoms are displaced along
the conjugate direction $\{\bh\}$ by a finite increment $\lambda$, as
$\bR_i \rightarrow \bR_i+\lambda \bh_i$.  A line minimization of
$E_{AE}$ along the conjugate gradient direction to find $\lambda$ that
minimizes $E_{AE}$ is done by taking advantage of the fact that
$E_{AE}$ is a fourth-order polynomial that depends on only the
relative positions, $\bR_i-\bR_j$, of each atom:
\begin{eqnarray}
E_{AE}[\{\bR+\lambda {\bf h}\}]
&=& E_{AE}[\{\bR_i-\bR_j\}]\nonumber\\
&&+\lambda E^{(1)}[\{\bR_i-\bR_j\},\{\bh_i-\bh_j\}]
+\lambda^2 E^{(2)}[\{\bR_i-\bR_j\},\{\bh_i-\bh_j\}]\nonumber\\
&&+\lambda^3 E^{(3)}[\{\bR_i-\bR_j\},\{\bh_i-\bh_j\}]
+\lambda^4 E^{(4)}[\{\bR_i-\bR_j\},\{\bh_i-\bh_j\}].
\label{ecg1}
\end{eqnarray}
The energy terms, $E^{(1)}, E^{(2)}, E^{(3)}$ and $E^{(4)}$,
are also simple polynomials of $\{\bR\}$ and $\{{\bf h}\}$.
The increment $\lambda$ minimizing the elastic energy
is then obtained by solving  exactly
\begin{equation}
{\partial E_{AE}[\{\bR+\lambda {\bf h}\}]\over \partial \lambda}
= E^{(1)}+2\lambda E^{(2)}+3\lambda^2 E^{(3)}
+4\lambda^3 E^4 = 0.
\label{ecg2}
\end{equation}

Figure~2 illustrates how the local strain is calculated.
After the atomic positions are relaxed by minimizing $E_{AE}$,
the local strain tensor $\tilde\eps$ at a cation site
(for cation-mixed systems)
is calculated by considering
a tetrahedron formed by four nearest neighboring anions.
The distorted tetrahedron edges, $\bR_{12},\bR_{23}$ and $\bR_{34}$,
are related to the
ideal tetrahedron edges $\bR_{12}^0,\bR_{23}^0$ and $\bR_{34}^0$
via
\begin{equation}
\left(\begin{array}{c}R_{12,x}\\R_{12,y}\\R_{12,z}\end{array}
\begin{array}{c}R_{23,x}\\R_{23,y}\\R_{23,z}\end{array}
\begin{array}{c}R_{34,x}\\R_{34,y}\\R_{34,z}\end{array}\right) =
\left(\begin{array}{c}1+\eps_{xx}\\\eps_{xy}\\\eps_{xz}\end{array}
\begin{array}{c}\eps_{yx}\\1+\eps_{yy}\\\eps_{yz}\end{array}
\begin{array}{c}\eps_{zx}\\\eps_{zy}\\1+\eps_{zz}\end{array}\right)
\left(\begin{array}{c}R_{12,x}^0\\R_{12,y}^0\\R_{12,z}^0\end{array}
\begin{array}{c}R_{23,x}^0\\R_{23,y}^0\\R_{23,z}^0\end{array}
\begin{array}{c}R_{34,x}^0\\R_{34,y}^0\\R_{34,z}^0\end{array}\right).
\label {EPS}
\end{equation}
The ideal tetrahedron edges
are $\{\bR^0\} = \{[110]a/2,[0\bar11]a/2,[\bar110]a/2\}$,
where $a$ denotes the equilibrium lattice constant of the cation, i.e.,
$a_{\rm GaAs}$ for Ga atoms
and $a_{\rm InAs}$ for In atoms.
The local strain, $\tilde\eps$, is then calculated
by a matrix inversion as
\begin{equation}
\left(\begin{array}{c}\eps_{xx}\\\eps_{xy}\\\eps_{xz}\end{array}
\begin{array}{c}\eps_{yx}\\\eps_{yy}\\\eps_{yz}\end{array}
\begin{array}{c}\eps_{zx}\\\eps_{zy}\\\eps_{zz}\end{array}\right)
=
\left(\begin{array}{c}R_{12,x}\\R_{12,y}\\R_{12,z}\end{array}
\begin{array}{c}R_{23,x}\\R_{23,y}\\R_{23,z}\end{array}
\begin{array}{c}R_{34,x}\\R_{34,y}\\R_{34,z}\end{array}\right)
\left(\begin{array}{c}R_{12,x}^0\\R_{12,y}^0\\R_{12,z}^0\end{array}
\begin{array}{c}R_{23,x}^0\\R_{23,y}^0\\R_{23,z}^0\end{array}
\begin{array}{c}R_{34,x}^0\\R_{34,y}^0\\R_{34,z}^0\end{array}\right)^{-1}-I,
\label {EPSp}
\end{equation}
where $I$ is the unit matrix.

\section{Results}
\subsection{Comparison of strain profiles}

Figure~3 shows $\eps_{xx},\eps_{zz}$ and ${\rm Tr}(\eps) =
\eps_{xx}+\eps_{yy}+\eps_{zz}$ as obtained by continuum elasticity
(dashed lines) and by atomistic elasticity (solid lines) as a function
of the position from the pyramidal center along the $[110]$ direction
at a height $z=h/3$ from the base (see Fig.~1).  The corresponding
differences in strains,$\Delta\eps=\eps(CE)-\eps(AE)$, are given as
the solid lines in Figure~4.  We note that the grid points of the
continuum elasticity calculation are chosen to be commensurate with
the cation positions of the ideal GaAs Zincblend structure for
consistent comparisons of the two approaches.  The largest differences
occur around the interfaces between the dot and the cap.  A
significant discrepancy is also found inside the quantum dot where the
InAs experience large compressive strains: $\eps_{xx}$ of the
continuum elasticity is found to be more compressive than that of the
atomistic elasticity, while the $\eps_{zz}$ of the CE is more tensile.
A similar comparison is given in Fig.~5, but this time the position
vector is along the $Z=[001]$ direction, starting from the substrate,
going through the wetting layer into the pyramidal tip and then into
the capping layer.  Again, the discrepancy is largest around the
interfaces, while the strains in the barrier (GaAs substrate and
capping layer) agree within 0.5\%.

Figure 4 illustrates the extent to which the
continuum elasticity description misses the correct atomic symmetry.
In a pyramid made of Zincblend materials
on the $(001)$ substrates,
the $\{110\}$ and $\{\bar110\}$ facets are symmetrically inequivalent
(Fig.~1).
Indeed, the atomistic calculation produces different strains.
The dashed lines in Fig.~4
show the difference $\eps_{ij}^{AE}([110])-\eps_{ij}^{AE}([\bar110])$
for these two directions.
We see that the anisotropy is pronounced at the interfaces.
For the atomistic elasticity calculation, we construct
the pyramidal structure to have an In atom at the pyramidal tip.
This tip In atom has: (i) two-As atoms that belong to the InAs dot
and lie along the $[110]$ direction;
and (ii) the other two-As atoms that belong to the GaAs capping layer
and lie along the $[\bar110]$ direction.
Considering only the local strain of the tip atom,
one expects  larger compressive $\eps_{xx}$ and $\eps_{yy}$
along the $[110]$ direction
than along the  $[\bar110]$ direction,
based on the atomic configuration.
By the same token,
the atoms at the $\{110\}$ interfaces
experience larger compressive $\eps_{xx}$ and $\eps_{yy}$,
than those at the $\{\bar110\}$ interfaces,
for this particular choice of the pyramidal geometry.
In the continuum description,
$\eps_{ij}^{CE}([110])=\eps_{ij}^{CE}([\bar110])$
and this effect is missing.

\subsection{The origin of the differences -- a simple test case}

We know that the continuum and atomistic models, starting from the
same input elastic constants, must agree in the limit of small strain
and a large system.  To study the rate at which the two methods
diverge with increasing strain, we consider the simple case of biaxial
strain.  As Eqs.~(\ref{ST})--(\ref{RF}) show, for a 2-dimensional film
that is constrained on a $(001)$ substrate, continuum elasticity
predicts
\begin{equation}
{\eps_{\perp}\over \eps_{||}} \equiv
\left[ {c - a_{eq}\over a_s - a_{eq}}\right]_{CE} = -{2C_{12}\over C_{11}}.
\label{biaxa}
\end{equation}
Figure~6(a) compares this result with that obtained via atomistic
elasticity, as a function of the relative film/substrate mismatch
$\eps_{||} = (a_s-a_{eq})/a_{eq}$.  Similarly, for the $(110)$ strain,
\begin{equation}
{\eps_{\perp}\over \eps_{||}} =
-{C_{11}+3C_{12}-2C_{44}\over C_{11}+C_{12}+2C_{44}},
\label{biaxb}
\end{equation}
and the corresponding comparison of the continuum and atomistic
elasticity is shown in Fig.~6(b).  We see that the discrepancy rises
linearly, reaching 4 \%\ for a lattice mismatch of 7 \%,
characteristic of InAs/GaAs.  This difference is comparable to that
found between CE and AE around the interfaces of the quantum dots
(Figs.~3--5).  Thus, the discrepancy simply reflects the departure
from the linearity regime of the continuum elasticity.

\subsection{Consequences of the different strains
in continuum elasticity and atomistic elasticity}

The existence of different
strain magnitudes and even symmetries
in a continuum elasticity
{\it vs.} atomistic elasticity descriptions
can affect the ensuing
electronic structure of the quantum dot.
Most notably, the real point group symmetry of
the square pyramid is $C_2$, but
continuum elasticity  spuriously  produces
 a higher $C_4$ symmetry.

Regarding the quantitative effects, there are different levels of
approximation for coupling the strain to the electronic structure.
The most general and accurate electronic structure approach is
atomistic ({\it e.g.}, pseudopotentials, tight-binding).  There, the
full set of atomic positions affects the electronic structure.  In
more approximate electronic structure approaches, such as continuum
effective-mass, only some aspects of the full,
position-dependent-strain tensor, $\eps_{ij}(\br)$ is ``felt'' by the
electronic structure.  In these approaches, one considers
strain-modified potential wells as barriers.  Since experiments
typically measure electronic energies rather than strains, it is
instructive to examine these effects.

Assuming decoupled conduction and valence bands
the strain-modified confinement potential of the conduction-band state
is
\begin{equation}
E_{c} (\br) = E_{c}^0(\br) + a_c (\br) {\rm Tr}[\eps(\br)].
\label{elCP}
\end{equation}
Here, $E_{c}^0(\br)$ is the energy of the conduction-band minimum
of the bulk material at \br\
and $a_c$, the deformation potential of the conduction-band
under hydrostatic deformation.
The ``strain'' Hamiltonian of the valence states~\cite{wei} is
\begin{eqnarray}
H_{v} &=& a_v(\br) {\rm Tr}(\eps(\br))
-b \left[\left(\begin{array}{c}-2\\0\\0\end{array}
\begin{array}{c}0\\1\\0\end{array}
\begin{array}{c}0\\0\\1\end{array}\right)\eps_{xx}
+\left(\begin{array}{c}1\\0\\0\end{array}
\begin{array}{c}0\\-2\\0\end{array}
\begin{array}{c}0\\0\\1\end{array}\right)\eps_{yy}
+\left(\begin{array}{c}1\\0\\0\end{array}
\begin{array}{c}0\\1\\0\end{array}
\begin{array}{c}0\\0\\-2\end{array}\right)\eps_{zz}
\right]\nonumber\\
&&-\sqrt{3}d
\left[\left(\begin{array}{c}0\\-1\\0\end{array}
\begin{array}{c}-1\\0\\0\end{array}
\begin{array}{c}0\\0\\0\end{array}\right)\eps_{xy}
+\left(\begin{array}{c}0\\0\\0\end{array}
\begin{array}{c}0\\0\\-1\end{array}
\begin{array}{c}0\\-1\\0\end{array}\right)\eps_{yz}
+\left(\begin{array}{c}0\\0\\-1\end{array}
\begin{array}{c}0\\0\\0\end{array}
\begin{array}{c}-1\\0\\0\end{array}\right)\eps_{zx}
\right],
\label{hH}
\end{eqnarray}
where $a_v$ is the hydrostatic deformation of the valence states
and $b$ and $d$ are uniaxial deformation potentials
for $(001)$ strain and $(111)$ strain, respectively.
The effective confinement potentials  of the valence
states are obtained by diagonalizing the strain Hamiltonian
coupled with the spin-orbit Hamiltonian~\cite{wei}.
Along the $[001]$ direction ($z-$axis) through the pyramidal tip,
the shear strains (off-diagonal terms of the strain tensor)
are zero and $\eps_{xx}=\eps_{yy}$, and thus
the effective confinement potentials can be simplified as
\begin{eqnarray}
E_{hh} &=& E_v^0 + a_v {\rm Tr}(\eps)+{1\over 3}[\Delta^{so}+\Delta^s(\eps)]
\nonumber\\
E_{lh} &=& E_v^0 + a_v {\rm Tr}(\eps)-{1\over 6}[\Delta^{so}+\Delta^s(\eps)]
+{1\over 2}\sqrt{(\Delta^{so}+\Delta^s(\eps))^2-
{8\over 3}\Delta^{so}\Delta^s(\eps)},
\label{holeCP}
\end{eqnarray}
where $\Delta^{so}$ is the spin-orbit splitting
and $\Delta^s \equiv -3b[\eps_{zz}-(\eps_{xx}+\eps_{yy})/2]$.

Figure~7 shows the effective confinement potentials of the conduction
and valence-band states along the $z-$axis through the tip of the
pyramid.  The strain profiles obtained by the continuum elasticity and
atomistic elasticity are used for the calculation with the same
material parameters given in Table II~\cite{bimberg95:1}.  Again, the
largest difference in confinement potentials is found at the
interfaces at about 100 meV for the conduction band and 200meV for the
valence band.  The average difference of the confinement potentials
inside the dot is about 20 meV for the conduction-band state.
Although the differences in the strain-modified-confinement potentials
are small, the band edge states are expected to show different
characteristics depending upon which strain profile is used for the
electronic structure calculation.

\section{Conclusions}

We compare the strain distribution of the pyramidal InAs dot grown on
a GaAs substrate calculated using continuum elasticity and atomistic
elasticity.  We find a significant difference in the strain around the
dot interfaces and inside the dot, while the difference in the barrier
(GaAs substrate and capping layer) is very small.  The difference
between the two results is attributed to the large strain outside the
linearity regime of CE, and to the loss of the correct atomic symmetry
by the CE.

\noindent {\bf Acknowledgements}
This work was supported by
United States Department of Energy -- Basic Energy Sciences, Division
of Materials Science under contract No. DE-AC36-83CH10093.

\newpage
\begin{table}[hbt]
\caption{Ideal bond lengths ($d^0$),
elastic constants, and force constants ($\alpha$ and $\beta$)
of bulk GaAs and InAs.
Elastic constants of valence force field method
are evaluated by Eq.~(8)
using $\alpha$ and $\beta$ given below.
}
\begin{tabular}{lcccccc}
& $d_{ij}^0$ & {$C_{11}$} & $C_{12}$ & $C_{44}$ & $\alpha$ & $\beta$\\
Material & (\AA) & \multicolumn{3}{c}{$10^{11}$dyne/cm$^{-1}$}
&\multicolumn{2}{c}{$10^{3}$dyne}  \\
\hline
GaAs (valence force field) & 2.448 & 12.03 & 5.70 &  5.20 & 41.49 & 8.94\\
GaAs (experimental) & 2.448 & 12.11 & 5.48 &  6.04 &   &  \\
InAs (valence force field) & 2.622 & 8.53 & 4.90 &  3.14 & 35.18 & 5.49\\
InAs (experimental) & 2.622 & 8.329 & 4.526 &  3.959 &  & \\
\end{tabular}
\end{table}
\newpage
\begin{table}[hbt]
\caption{Material parameters used for Fig.~7.
All the numbers are given in eV from Ref.~[2].
}
\begin{tabular}{lcccccc}
& $\Delta^{so}$ & $V_{v}^0$ & $a_v$ & b & $V_{c}^0$ & $a_c$ \\
\hline
GaAs & 0.34 & 0 & 1.16 &  -1.6 & 1.52 & -8.33\\
InAs & 0.38 & 0.25 & 1.00 &  -1.8 & 0.66 & -6.08\\
\end{tabular}
\end{table}

\newpage
\begin{figure}
\caption{Schematic diagram of a square pyramidal InAs dot on
$(001)$ GaAs substrate.  The wetting layer (WL) consists of 1
monolayer (ML) of In atoms.  Three principal directions, $[100],[010]$
and $[001]$ are denoted as $X, Y$, and $Z$.  The orientation of the
pyramidal base is $X\times Y$ and the ratio of the base length ($b$)
and the height ($h$) is 2 with $\{110\}$ (grey) and $\{\bar 110\}$
facets.  Although not shown in the figure, the pyramidal InAs dot is
capped by GaAs.  }
\end{figure}
\begin{figure}
\caption{Schematics to illustrate how the local strain
is calculated.
For a cation (Ga or In), three vectors ($\{\bR\}$)
forming a distorted tetrahedron
after atomic relaxation
are related to the equivalent vectors ($\{\bR^0\}$)
of an ideal tetrahedron via the strain tensor.
}
\end{figure}

\begin{figure}
\caption{Strain profiles along the $[110]$ direction
at $z = h/3$ from the base of the pyramid.
The solid lines are the strain  profiles obtained by atomistic elasticity
and dotted lines those by continuum elasticity.
The positive and negative signs of $X-$axes denote the $[110]$
and $[\bar1\bar10]$ direction, respectively.
}
\end{figure}

\begin{figure}
\caption{Solid lines denote
$\Delta\eps_{ij} \equiv \eps_{ij}({\rm CE})-\eps_{ij}({\rm AE})$,
the difference of each strain component
obtained by the continuum elasticity and the atomistic elasticity calculations.
In III-V Zincblend semiconductors,
the $[110]$ and $[\bar110]$ directions
are inequivalent and, therefore, the symmetry of the pyramid is $C_2$.
The $C_2$ symmetry is apparently seen by
the difference of the strains (dashed lines)
along the $[110]$ and $[\bar110]$ directions (see Fig.~1)
}
\end{figure}

\begin{figure}
\caption{Strain profiles and the differences
along the $Z$ direction through the pyramidal tip.
The differences between the CE and AE are
given on the right-hand side.
The discrepancy is largest around the interfaces,
while the strains in the barrier (GaAs substrate and capping layer)
agree well within 0.5\%.
A significant difference is also found inside
 the quantum dot where
the InAs experience large compressive strains at about 7 \%\
due to the lattice mismatch.
}
\end{figure}

\begin{figure}
\caption{Relaxation of $c_{eq}$ of GaAs under a biaxial strain
obtained by atomistic elasticity
is compared to the prediction of the continuum elasticity:
(a) $\bG = [001]$ and $\eps_{\perp}/\eps_{||} = -2C_{12}/C_{11}$;
and (b) $\bG = [110]$ and
$\eps_{\perp}/\eps_{||} = -(C_{11}+3C_{12}-2C_{44})/(C_{11}+C_{12}+2C_{44})$.
At the infinitely small limit,
the $-\eps_{\perp}/\eps_{||}$ of the atomistic elasticity (solid lines)
coincides with that of the continuum elasticity (dashed lines).
The discrepancy between the CE and AE increases
for the larger biaxial strains.
}
\end{figure}

\begin{figure}
\caption{(a) Confinement potentials by Eq.~(15) and (16),
along $Z$ direction through the pyramidal tip in Fig.~1,
predicted by the strain profiles obtained by
the atomistic elasticity.
All the energies are measured with respect to
the valence band maximum of the bulk GaAs at equilibrium.
(b) The difference of the confinement potentials by continuum
elasticity and atomistic elasticity.}
\end{figure}

\end{document}